\documentclass{iopart}
\usepackage[utf8]{inputenc}
\usepackage{bm}
\usepackage{graphicx}
\usepackage{enumerate}
\usepackage{indentfirst}
\usepackage{color}

\newcommand{\mody}{\color{black}}
\newcommand{\norm}{\color{black}} 

\begin{document}
\title{Analysing the impact of anisotropy pressure on tokamak equilibria}
\author{Z.S. Qu, M. Fitzgerald and M.J. Hole}
\address{Research School of Physics and Engineering, the Australian National University, Canberra ACT 0200, Australia}

\begin{abstract}
Neutral beam injection or ion cyclotron resonance heating induces pressure anisotropy. 
The axisymmetric plasma equilibrium code HELENA has been upgraded to include anisotropy and toroidal flow. 
With both analytical and numerical methods, 
we have studied the determinant factors in anisotropic equilibria and their impact on flux surfaces, magnetic axis shift, 
the displacement of pressures and density contours from flux surface. 
With $p_\parallel/p_\perp \approx 1.5$, $p_\perp$ can vary 20\% on $s=0.5$ flux surface, in a MAST like equilibrium. 
We have also re-evaluated the widely applied approximation to anisotropy in which $p^*=(p_\parallel + p_\perp)/2$, 
the average of parallel and perpendicular pressure, is taken as the approximate isotropic pressure. 
We show that an isotropic reconstruction can infer a correct $p^*$,
only by getting an incorrect $RB_\varphi$.
\mody We find the reconstructions of the same MAST discharge with $p_\parallel/p_\perp \approx 1.25$,
using isotropic and anisotropic model respectively,
to have a 3\% difference in toroidal field but a 66\% difference in poloidal current. \norm
\end{abstract}

\section{Introduction} \label{sec:introduction}
Auxiliary heatings, such as neutral beam injection (NBI) and ion cyclotron resonance heating (ICRH), are widely implemented in modern tokamaks. 
Unlike Ohmic heating, NBI and ICRH generate a large population of fast ions.
The NBI induced energetic ions mainly come with a large energy parallel to injection, 
while ICRH heats the ions into higher velocities perpendicular to magnetic field.\cite{review-nf-2007} 
The distribution functions of these fast ions in phase space are thus distorted into anisotropic forms with $p_\perp \neq p_\parallel$,
where $p_\perp$ or $p_\parallel$ refers to the total pressure of both the thermal and the fast population perpendicular or parallel to the magnetic field.
These heating methods also drive plasma rotation.
The resulting magnitude of anisotropy in a tokamak can be very large according to recent studies.
In JET, anisotropy magnitude reaches $p_\perp / p_\parallel \approx 2.5$ \cite{zwingmann-ppcf-2001} with ICRH.
In MAST, the beam pressure reaches $p_\perp / p_\parallel \approx 1.7$ during NBI heating \cite{hole-ppcf-2011}. 

\mody However, in the magnetohydrodynamic(MHD) description of plasma, 
pressure is assumed to be isotropic.
Three questions are raised immediately. 
How is an anisotropic equilibrium different from an isotropic one?
How accurate is the MHD model for anisotropic equilibria?
How does the change in equilibrium affect the further study of a plasma (such as stability and transport)? \norm

The theory of tokamak anisotropic equilibrium has been studied by many authors
\cite{northrop-prl-1964, grad-pf-1967,dobrott-pf-1970,spies-pf-1974,salberta-pf-1987}.
One basic result is that the two pressures $p_{\parallel,\perp}$ and the density $\rho$ are no longer flux functions \cite{cooper-nf-1980, iacono-pfb-1990, pustovitov-ppcf-2010}. 
At the same time, anisotropy could add to or subtract the magnetic axis outward shift (Shafranov-shift \cite{shafranov-ropp-1986}) \cite{iacono-pfb-1990, madden-nf-1994, pustovitov-aipcp-2012}.
The latter result has been confirmed by numerical code FLOW \cite{flow-pop-2004}.
Some authors also find the experimentally inferred equilibrium assuming single pressure and anisotropic pressure can be quite different 
\cite{zwingmann-ppcf-2001, hole-ppcf-2011, hole-ppcf-2013}.

\mody In this work, we address the first two questions with analytical and numerical approaches. \norm
We show how $p_\parallel$, $p_\perp$ and the ``nonlinear'' part separately contribute to the force balance and deviate from flux functions.
We also answer the second question of what problem a scalar pressure model will lead to in equilibrium reconstruction, 
and its dependency on aspect ratio and the magnitude of anisotropy.

This work is organized as follows:
In Section \ref{sec:gseq}, the anisotropic and toroidal flowing modified Grad-Shafranov equation we use in our analytical and numerical study is derived and presented.
Section \ref{sec:numerical} briefly describes the numerical methods and the code HELENA+ATF.
The features of an anisotropic equilibrium are studied in Section \ref{sec:ani-characteristics}.
Section \ref{sec:mhd-approximation} evaluates the widely used MHD scalar pressure approximation to anisotropic pressure.

\section{Grad-Shafranov Equation with anisotropic pressure and toroidal flow}\label{sec:gseq}
\subsection{Basic Equations}
Our assumptions of plasma equilibrium are based on guiding center plasma theory (GCP) \cite{dobrott-pf-1970,grad-guiding-1967} with ideal MHD Ohm's law.
The basic equations are (in S.I. units): 
\begin{equation}
\rho(\bm{u} \cdot \nabla \bm{u}) + \nabla \cdot {P} = \bm{J} \times \bm{B},
\label{eq:basic-momentum}
\end{equation}
\begin{equation}
\nabla \times \bm{B} = \mu_0 \bm{J},
\label{eq:basic-maxwell1}
\end{equation}
\begin{equation}
\nabla \cdot \bm{B} = 0,
\label{eq:basic-maxwell2}
\end{equation}
\begin{equation}
\nabla \times \bm{E} = 0,
\label{eq:basic-maxwell3}
\end{equation}
\begin{equation}
 \bm{E} + \bm{u} \times \bm{B} = 0,
 \label{eq:basic-ohm}
\end{equation}
\begin{equation}
 {P} = p_{\perp} {I} + \frac{\Delta}{\mu_0} \bm{B} \bm{B}, \ \
\Delta \equiv \mu_0  \frac{p_{\parallel} - p_{\perp}} {B^2},
\label{eq:basic-gcp-pressure}
\end{equation}
where $\rho$ is the mass density, $\bm{u}$ the single fluid velocity, 
${P}$ the pressure tensor, $\bm{J}$ the current density,
$\bm{B}$ the magnetic field, $\bm{E}$ the electric field, and $\mu_0$ the vacuum permeability constant.
Equation (\ref{eq:basic-momentum}) is the GCP force balance.
Equation (\ref{eq:basic-maxwell1}), (\ref{eq:basic-maxwell2}) and (\ref{eq:basic-maxwell3}) are Maxwell equations.
Equation (\ref{eq:basic-ohm}) is the ideal Ohm's law.
Equation (\ref{eq:basic-gcp-pressure}) is the GCP assumption of anisotropic pressure,
which assumes the pressure tensor consists of two components, 
$p_\perp$ and $p_\parallel$,
with $I$ the identity tensor.
\mody The fast ion finite orbit width (FOW) effects are ignored in our fluid model.
 FOW effects can be important for tokamaks with fast ion heating, especially in tight aspect ratio tokamaks.
 For example, the fast ion orbit width can be as large as $20\%$ of the minor radius in MAST with parallel on-axis beam.
The inclusion of these effects in equilibrium requires a kinetic/gyro-kinetic treatment of the fast ions
(e.g. the inclusion in fast ion currents and thus the equilibrium, when fast ion proportion is low \cite{belova-pop-2003, todo-pop-2005}). \norm

With axisymmetric cylindrical coordinate system $(R,Z,\varphi)$ and Eq. (\ref{eq:basic-maxwell2}),
$\bm{B}$ is written as
\begin{equation}
 \bm{B} = \nabla \Psi \times \nabla \varphi + R B_\varphi \nabla \varphi,
 \label{eq:b-definition}
\end{equation}
where $\Psi$ is the poloidal magnetic flux and $B_\varphi$ the toroidal magnetic field.
Current density in toroidal and poloidal direction can be deduced from the Ampere's Law (Eq. (\ref{eq:basic-maxwell1})) :
\begin{equation}
 \mu_0 J_\varphi = -R \nabla \cdot \frac{ \nabla \Psi} {R^2}, \ \ 
 \mu_0 \bm{J_p}  = \nabla (RB_\varphi) \times \nabla \varphi.
 \label{eq:jphi}
\end{equation}

If only the toroidal part of flow is important, 
with $\nabla \times (\bm{u} \times \bm{B})=0$ from Eq. (\ref{eq:basic-maxwell3}) and (\ref{eq:basic-ohm}),
the form of $\bm{u}$ becomes
\begin{equation}
 \bm{u} = \Omega(\Psi) R^2 \nabla \varphi,
\end{equation}
in which $\Omega$ is the toroidal angular velocity and a flux function for zero resistivity.

Two different forms of toroidal flow and anisotropic modified Grad-Shafranov equations (modified GSE) \cite{grad-pf-1967,shafranov-ropp-1986} can be derived from the above equations using different variables.
The pressure form of the GSE has pressures as a function of three variables $(R,B,\Psi)$ : $p_{\parallel,\perp} = p_{\parallel,\perp} (R,B,\Psi)$ 
\cite{salberta-pf-1987, cooper-nf-1980, pustovitov-aipcp-2012, grad-guiding-1967, fitzgerals-subnf-2013}.
The enthalpy form uses $\rho$ as a variable instead of $R$, which means $p_{\parallel,\perp} = p_{\parallel,\perp} (\rho,B,\Psi)$ 
\cite{iacono-pfb-1990, flow-pop-2004,fitzgerals-subnf-2013}. 

\subsection{Grad-Shafranov Equation in the form of pressure}
To obtain the modified GSE in the pressure form, 
the momentum equation is rearranged into a form, as mentioned by many authors 
(for example \cite{northrop-prl-1964, grad-pf-1967,iacono-pfb-1990,pustovitov-ppcf-2010, flow-pop-2004, fitzgerals-subnf-2013}) :
\mody \begin{equation}
 \mu_0 \nabla p_\parallel = \Delta \nabla \frac{B^2}{2} + \nabla \times [(1-\Delta) \bm{B}] \times \bm{B} + \mu_0 \rho \Omega^2 R \nabla R.
 \label{eq:force-another-form}
\end{equation} \norm
Substituting $p_{\parallel} = p_{\parallel} (R,B,\Psi)$ into Eq. (\ref{eq:force-another-form}),
the component of Eq. (\ref{eq:force-another-form}) in $\nabla \varphi$, $\nabla B$, $\nabla R$ and $\nabla \Psi$ directions each gives
\begin{equation}
 F(\Psi) \equiv RB_\varphi(1-\Delta) ,
 \label{eq:f-definition}
\end{equation}
\begin{equation} 
 \left( \frac{\partial p_\parallel}{\partial B} \right) _{\Psi, R} = \frac{\Delta B} {\mu_0},
 \label{eq:gs-restrictions}
\end{equation}
\begin{equation}
 \left( \frac{\partial p_\parallel}{\partial R} \right) _{\Psi, B} = \rho R \Omega^2,
 \label{eq:gs-restrictions-r}
\end{equation}
\begin{equation}
\nabla \cdot \frac{(1-\Delta) \nabla \Psi} {R^2} = -\frac{FF^{\prime}}{(1-\Delta)R^2} - \mu_0 \left( \frac{\partial p_\parallel}{\partial \Psi} \right) _{R,B}.
\label{eq:gs-2}
\end{equation}
We note that $F=RB_\varphi(1-\Delta)$, instead of $RB_\varphi$, becomes a flux function.
The restrictions for $p_\parallel(R,B,\Psi)$ are Eq. (\ref{eq:gs-restrictions}) and (\ref{eq:gs-restrictions-r}):
these also guarantee the parallel force balance (multiplying Eq. (\ref{eq:force-another-form}) by $\bm{B}$) is satisfied.
In the limit of no toroidal flow, Eq. (\ref{eq:gs-restrictions}) can also be deduced from the parallel force balance.
Finally, Eq. (\ref{eq:gs-2}) is the modified GSE for anisotropic and toroidally rotating system.

\subsection{Grad-Shafranov Equation in the form of enthalpy}
\mody A detailed derivation of the enthalpy form of the modified GSE can be found in \cite{iacono-pfb-1990, flow-pop-2004, fitzgerals-subnf-2013}.
Starting from the energy conservation equation,
the relationships between the enthalpy $W(\rho,B,\Psi)$ and plasma pressures as well as rotation are derived.
A new flux function $H$, which defines as
\begin{equation}
 H(\Psi) = W(\rho,B,\Psi) - \frac{1}{2} \Omega^2 R^2,
\end{equation}
is inferred from these relationships. \norm

\mody In order to close the set of equations, a certain equation of state is needed.
In our work, the bi-Maxwellian distribution model is chosen.
This is the simplest distribution function that will capture anisotropy.
The two pressures are now products of plasma density and the parallel and perpendicular temperatures,
and the thermal closure chosen is that parallel temperature is a flux function: \norm
\begin{equation}
p_{\parallel}(\rho,B,\Psi) = \rho T_{\parallel}(\Psi), \ 
p_{\perp}(\rho,B,\Psi) = \rho T_{\perp}(B,\Psi).
\label{eq:bi-maxwell-p}
\end{equation}
The two temperatures $T_{\parallel}$ and $T_{\perp}$ are in units of energy per mass.
Inserting the bi-Maxwellian assumptions yields a expression for $W(\rho, B, \Psi)$ and $T_{\perp}(B,\Psi)$
\cite{iacono-pfb-1990,fitzgerals-subnf-2013},
written as
\begin{equation}
 W(\rho, B, \Psi) = T_{\parallel} \ln{\frac {T_{\parallel} \rho} {T_{\perp} \rho_0}}, \ \ \ 
 \rho = \rho_0 \frac{T_{\perp}}{T_{\parallel}}\exp{\frac{H + \frac{1}{2} R^2 \Omega^2} {T_{\parallel}}},
 \label{eq:enthalpy-density-form}
\end{equation}
\begin{equation}
T_{\parallel} = T_{\parallel}(\Psi),\ T_{\perp} = \frac{T_{\parallel}B} {|B - T_{\parallel} \Theta(\Psi)|}, 
\label{eq:bi-maxwell}
\end{equation}
\mody with $\rho_0$ a constant and a new flux function $\Theta$ indicating the magnitude of anisotropy.

Considering the $\nabla \Psi$ direction of Eq. (\ref{eq:force-another-form}) will give the enthalpy form of the modified GSE : \norm
\begin{eqnarray}
\fl \nabla \cdot \frac{(1-\Delta) \nabla \Psi} {R^2} = 
-\frac{FF^{\prime}}{(1-\Delta)R^2} \nonumber\\
- \mu_0 \rho \left[ T_\parallel' + H' + R^2 \Omega \Omega' - \left(\frac {\partial W} {\partial \Psi}\right)_{\rho, B}\right],
\label{eq:gs}
\end{eqnarray}
with $F$ defined by Eq. (\ref{eq:f-definition}).
The system is specified by five functions $\{T_\parallel, H, \Omega, F, \Theta \}$ of $\Psi$ and the boundary conditions on $\Psi$.

The pressure form of the modified GSE (Eq. (\ref{eq:gs-2})), when closed with Eq. (\ref{eq:bi-maxwell}),
is equivalent to the enthalpy form of the modified GSE. The enthalpy form of the modified GSE with bi-Maxwellian assumption is numerically solved.
We have used the pressure form of the modified GSE to explore physics of anisotropic plasma.

\section{Numerical scheme} \label{sec:numerical}
Based on the modified GSE in Eq. (\ref{eq:gs}), 
we altered and updated the axisymmetric plasma equilibrium code HELENA \cite{helena-ccp-1991} to its anisotropy and toroidal flow version HELENA+ATF.
Since the internal physical assumptions and equations are completely changed, 
we have rewritten most of its matrix element calculations and post-processing, 
but have retained subroutines for isoparametric meshing. 
HELENA+ATF uses the same isoparametric bicubic Hermite elements as HELENA \cite{helena-ccp-1991, goedbloed-am-2010}.

\mody Equation (\ref{eq:gs}) is solved in its weak form. 
That is, with the spatial discretization in Ref. \cite{helena-ccp-1991} and \cite{goedbloed-am-2010},
the PDE system is transformed into a linear algebra problem by integrating both sides after multiplied by each Hermite element.
Here, a Picard iteration is used to solve the system.
The flux functions and $\Delta$ of $n$'th iteration is used to calculate the flux surfaces $\Psi(R,Z)$ of $(n+1)$'th iteration. \norm

If $p_\parallel>p_\perp$, $1-\Delta$ can go from positive to negative.
In this case, the shear Alfv\'{e}n wave becomes purely growing \cite{tajiri-jpsj-1967}, labeled as the firehose instability.
On the other hand, if $p_\parallel<p_\perp$, 
the mirror instability may occur, 
with the non-oscillating mode becoming unstable \cite{tajiri-jpsj-1967}.
The firehose and mirror stability criteria given by \cite{grad-pf-1967,parker-pr-1958} are
\begin{equation}
 1-\Delta > 0,
 \label{eq:mirror}
\end{equation}
\begin{equation}
 1 + \frac{\mu_0}{B} \frac{\partial p_\perp}{\partial B} > 0,
 \label{eq:firehose}
\end{equation}
which guarantee Eq. (\ref{eq:gs}) to be elliptic all the time \cite{iacono-pfb-1990, grad-guiding-1967}.
These criteria are also sufficient conditions for the solvability (see \ref{app-solve}) of the four interdependent variables 
$p_\parallel, p_\perp, B$ and $\Delta$ (Eq. (\ref{eq:basic-gcp-pressure}) (\ref{eq:b-definition}) and (\ref{eq:f-definition})).
In this work, we only discuss equilibria within these stability criteria.
With bi-Maxwellian Eq. (\ref{eq:bi-maxwell}), the stability criteria are written as
\begin{equation}
  \frac{3\beta_E + 2 + \sqrt{(3\beta_E + 2)^2 + 12\beta_E}} {6\beta_E} > \frac{p_\perp}{p_\parallel} > \frac{3\beta_E-2}{3\beta_E+4},
\end{equation}
with $\beta_E = \mu_0 (4p_\perp/3 + 2p_\parallel/3)/B^2$ the local ratio of the kinetic energy to the magnetic energy.
Even in a tokamak with $\beta_E=0.4$,
we still have the upper limit 3 and lower limit below zero.
Therefore, these stability criteria are satisfied in most scenarios,
although the mirror instability criterion may be approached in high $\beta$ tokamaks with strong ICRH or perpendicular NBI heating.

In order to benchmark force balance convergence of HELENA+ATF, 
we consider a test case with constant $F$ and $\Theta$ profiles, linear $T_\parallel$ profile ($\sim 1-\Psi$), and quadratic $H$ and $\Omega^2$ profiles($\sim (1-\Psi)^2$).
The plasma boundary is set to have elongation $\kappa=1.2$, triangularity $\delta=0.2$ and inverse aspect ratio $\epsilon=0.3$.
In anisotropic test cases, $p_\parallel/p_\perp = 1.5$ on the axis,
while in test cases with toroidal flow, $\bar{\Omega}^2/\bar{T}_\parallel = 0.5$ on the axis.
\begin{figure}[!htbp]
  \centering
$  \begin{array}{c c}
   \includegraphics[width=6cm]{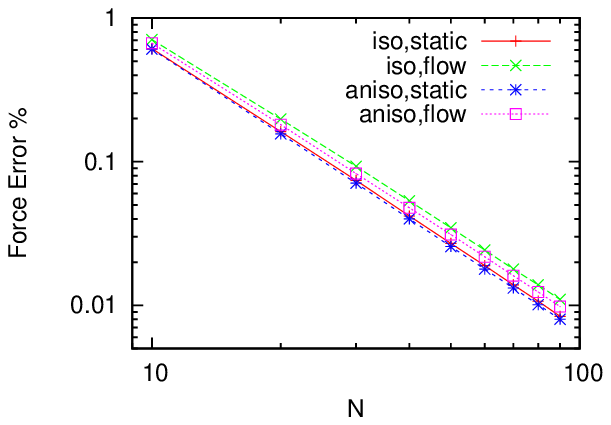} & \includegraphics[width=6cm]{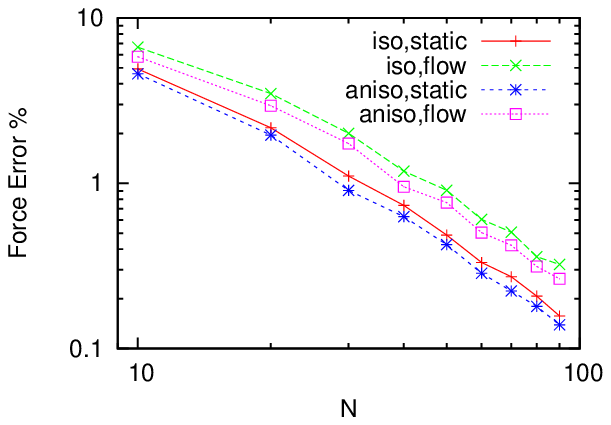} \\
   (a) & (b) \\
  \end{array}$
  \caption{(a) The average force balance error of all grid cells and (b) maximum force balance error in four test cases. 
  NR=NP=N is the number of radial and poloidal grid points. 
  Force balance error per cell means $\Delta F/F = 2|$RHS-LHS$|/|$RHS+LHS$|$ of Eq. (\ref{eq:gs}) in percent.
  Average force balance is calculated by $\sum (\Delta F/F)/N^2$.
  }
  \label{fig:numtest}
\end{figure}

Figure \ref{fig:numtest} shows the average force balance error of all grid cells and the maximum force balance error in four test cases.
The force balance error decreases logarithmically as grid resolution increases.
To explain the difference between Fig. \ref{fig:numtest} (a) and (b), we mention that
the force balance error is close to zero near the core but reach its maximum at the boundary.
This is not only because the grid is more concentrated at the core,
but also a sharp boundary approaching an X point or triangular point will cause numerical degrading with a singular Jacobian.

\mody Once the equilibrium is computed, HELENA+ATF also provides high precision coordinate information for stability codes. \norm
The solution of the modified GSE is mapped into the straight field line coordinate $(s,\vartheta, \varphi)$, which is defined as
\begin{equation}
s = \sqrt{\Psi/\Psi_0}, \ \ 
\vartheta(\theta) = \frac{F(\Psi)}{q}\int_\Psi \frac{dl} {R(1-\Delta)|\nabla \Psi|},
\end{equation}
where $q$ is defined as
\begin{equation}
 q(\Psi) = \frac{F(\Psi)}{2\pi}\oint_\Psi \frac{dl} {R(1-\Delta)|\nabla \Psi|}.
\end{equation}
The metric coefficients $g^{ij}$ and Jacobian $J$ can then be calculated.

\section{The features of anisotropic equilibria} \label{sec:ani-characteristics}
There are three major effects of anisotropic pressure that we can infer from our model and Eq. (\ref{eq:gs}):

\begin{enumerate}[(i)]
\item $p_\perp$ and $p_\parallel$ contribute separately to the toroidal current;
\item the term, ``$1-\Delta$'' inside the LHS operator will modulate the poloidal flux and form a new ``nonlinear current'';
\item pressures and density contours no longer lie on surfaces of constant poloidal flux.
\end{enumerate}

Effect (i) and (ii) will be explained in Section \ref{sec:flux-surface-shape}, and (iii) in Section \ref{sec:dev-flux-function}.
In this section, flow is turned off unless otherwise specified.
We choose profiles that represent the general shape and trend of the EFIT-TENSOR
reconstructed profiles with TRANSP\cite{transp-1979} constraint of MAST discharge \#18696 at 290ms \cite{fitzgerals-subnf-2013}.
They are
\begin{eqnarray}
  T(\Psi) = C_0(1-\Psi)^2 + C_1, \ \ 
  H(\Psi) = \frac{C_0}{2}(1-\Psi)^3 + C_2, \nonumber\\
  F(\Psi) = F_0, \ \
  \Theta(\Psi) = \Theta_0,
  \label{eq:profile-choice}
\end{eqnarray}
where $C_0, C_1, C_2, F_0$ and $\Theta_0$ are adjustable constants. 
Constants $C_1$ and $C_2$ are small values to make density and current profiles vanish at the plasma edge.
By varying $F_0$, we can adjust $q_0$.
The parameter $\Theta_0$ is associated with the magnitude of anisotropy.

For these profiles we examine four equilibrium configurations.
Equilibrium A is guided by a MAST like boundary with triangularity $\delta=0.4$, 
elongation $\kappa=1.7$ and inverse aspect ratio $\epsilon=0.7$.
Anisotropy of the case is chosen to be $p_\parallel / p_\perp \approx 1.5$ at core,
with a monotonic $q$ profile and $q_0\approx1$.
Equilibrium B examines the dependence with aspect ratio: $\epsilon$ is changed to $0.3$,
and $F_0$ adjusted to leave $q_0$ unchanged.
Equilibrium C examines the isotropic limit: $\Theta_0$ is set to zero,
and $F_0$ adjusted to leave $q_0$ unchanged.
Finally, equilibrium D examines the impact of toroidal flow,
with $\Omega^2 \sim (1-\Psi)^3$,
such that the ion thermal Mach number $M_{t\varphi}$ peaks at $0.7$ on axis and vanishes at the edge, 
where $M_{t\varphi} = v_\varphi / \sqrt{k_B T_i/ m_i}$ and $T_i$ is the ion temperature.
This is the typical upper limit of toroidal flow in MAST \cite{mast-ppcf-2003}.
In all cases anisotropy peaks at the core due to the flat $\Theta$ profile we have chosen.
Table \ref{tb:sample} shows parameters of these equilibria.

\begin{table}[!htbp]
\caption{\label{tb:sample}Parameters of equilibrium A, B, C and D.}
\begin{indented}
\item[]\begin{tabular}{@{}cccccc}
\br
  Equilibrium   & $\epsilon$ & $q_0$ &$\Delta$ & Anisotropy                          & Flow                         \\ \mr
  A             & 0.7        & 1.04  & $5.0\%$ & $p_\parallel / p_\perp \approx 1.5$ & none                         \\ 
  B             & 0.3        & 1.04  & $1.5\%$ & $p_\parallel / p_\perp \approx 1.5$ & none                         \\ 
  C             & 0.3        & 1.01  & $0.0\%$ & none                                & none                         \\ 
  D             & 0.7        & 1.05  & $0.0\%$ & none                                & $M_{t\varphi}\approx0.7$ on axis   \\ 
\br
\end{tabular}
\end{indented}
\end{table}

\subsection{Toroidal current decomposition} \label{sec:flux-surface-shape}
In a cylindrical plasma with straight field lines and infinite length, the perpendicular force balance is determined by $p_\perp$. 
In a tokamak, there is a $p_\parallel$ contribution \cite{pustovitov-aipcp-2012} to perpendicular force balance.
If flow is ignorable, we can rewrite Eq. (\ref{eq:gs-2}) and decompose $J_\varphi$ as
\begin{eqnarray}
\fl   \mu_0 J_\varphi = \underbrace{\mu_0 R \sin^2{\alpha}  \left( \frac{\partial p_\parallel}{\partial \Psi} \right)_{B}}_{J_{p_\parallel}} 
                &+ \underbrace{\mu_0 R \cos^2{\alpha}  \left( \frac{\partial p_\perp}{\partial \Psi} \right)_{B}}_{J_{p_\perp}}\nonumber\\
                &+ \underbrace{\frac{1-\Delta}{2 R} \left( \frac{\partial (RB_\varphi)^2 }{\partial \Psi} \right)_{B}}_{J_{tf} (toroidal field)}  
                 -\underbrace{R  \nabla \cdot \frac{ \Delta \nabla \Psi} {R^2}}_{J_{nl} (nonlinear)},
\label{eq:gs-jphi}
\end{eqnarray}
where $\alpha$ is the field pitch angle, i.e. $\tan \alpha \equiv B_p/B_\varphi$,
with $B_p$ the poloidal magnetic field.
The flux surface is determined by $J_\varphi$ through Eq. (\ref{eq:jphi}).
The four contributing terms, $J_{p_\parallel}, J_{p_\perp}$, $J_{tf}$ and $J_{nl}$ are identified here.
This equation shows that the balance of $J_{p_\perp}$ and $J_{p_\parallel}$ is determined by the pitch angle $\alpha$.

\begin{figure}[!htb]
\centering
$ \begin{array}{c c}
  \includegraphics[width=6cm]{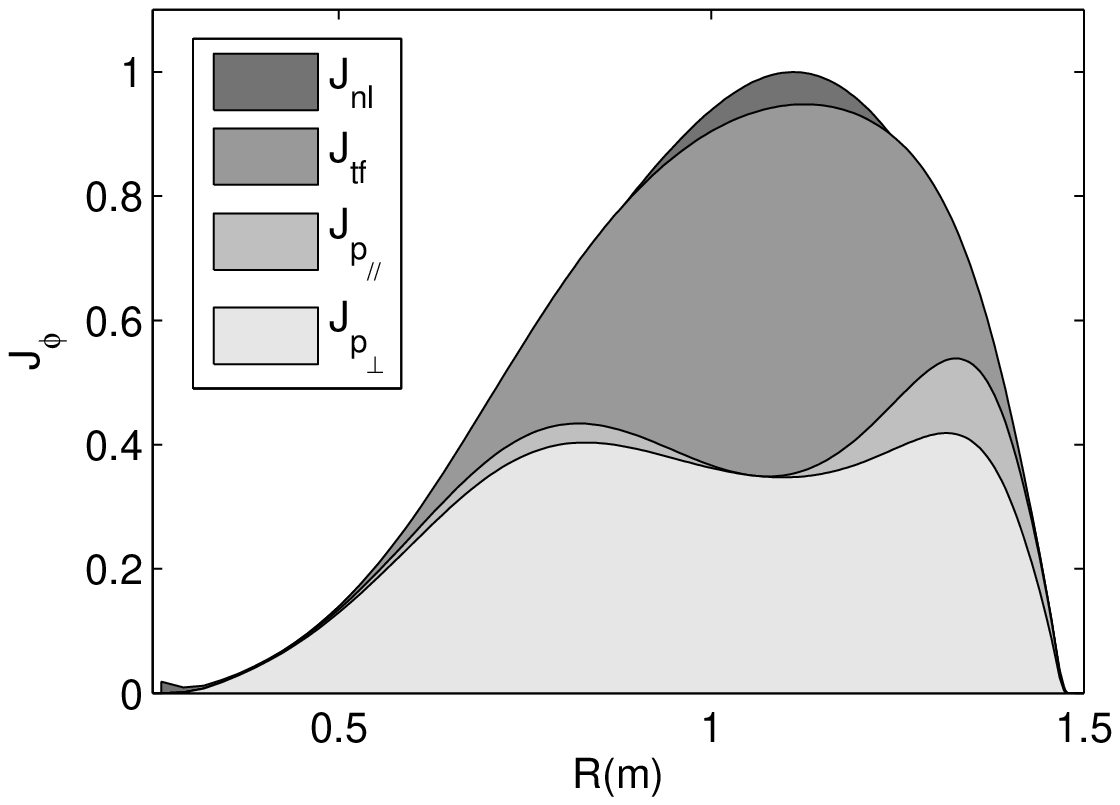} & \includegraphics[width=6cm]{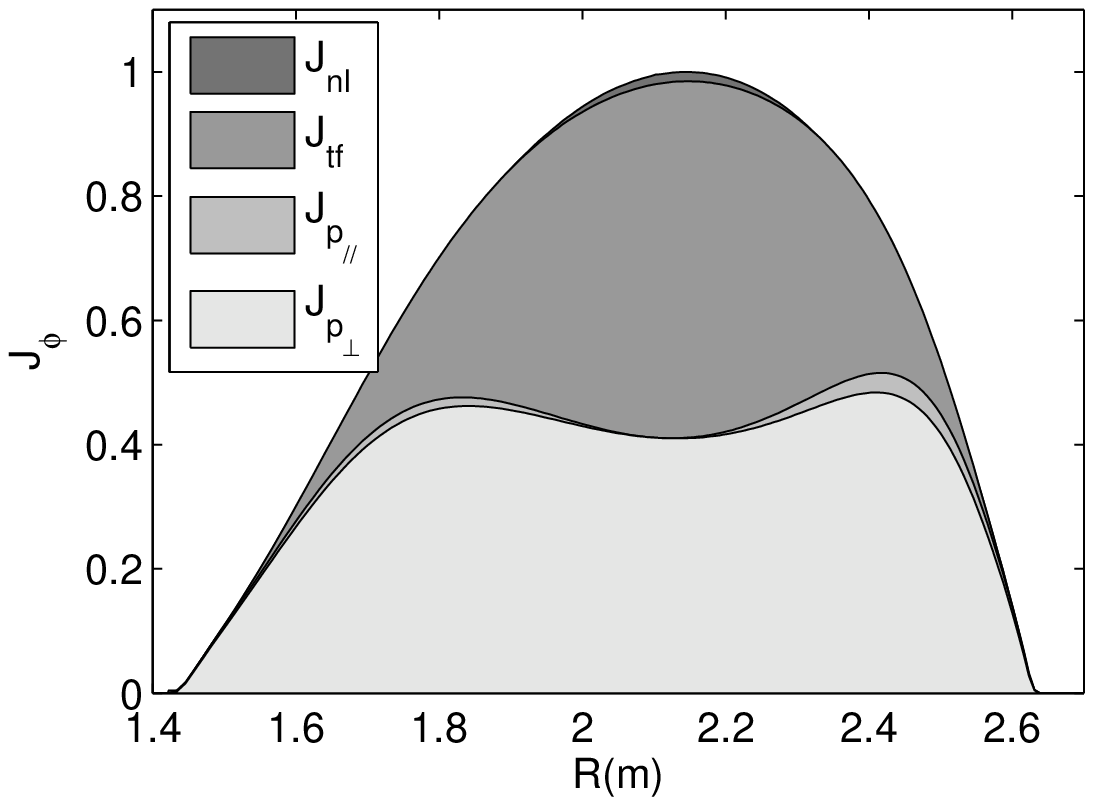} \\
  (a) & (b)
 \end{array}$
 \caption{Contribution of each component to $J_\varphi$ across the mid-plane in (a) equilibrium A with $\epsilon = 0.7$, 
          (b) equilibrium B with $\epsilon = 0.3$. 
          Shaded areas with different gray levels indicate different components.
          Maximum of $J_\varphi$ is normalized to unity.}
 \label{fig:jphi-contribution}
\end{figure}

Figure \ref{fig:jphi-contribution} shows the decomposition of $J_\varphi$ along the mid-plane for equilibrium A and B.
These two equilibria have similar profiles and their major difference is the aspect ratio.
In both cases, $J_\varphi$ is dominated by $J_{p_\perp}$ and $J_{tf}$, which roughly equal.
The $J_{p_\parallel}$  component is zero on the magnetic axis,
consistent with $\sin^2{\alpha} = B_p^2/B^2$ and $B_p = 0$ on axis.
For a low $\beta$ plasma, $\sin^2{\alpha} = B_p^2/B^2 \sim \epsilon^2 / q^2$.
We would thus expect, and observe, an increasing contribution from $J_{p_\parallel}$ with increasing $\epsilon$.
\mody For $\epsilon=0.7$,  $J_{p_\parallel}$ peaks at $20\%$ on the low field side.
Therefore, if the contribution of $p_\parallel$ is ignored, or in other words,
attributed to $p_\perp$, 
the current profile, and thus the $q$ profile will be changed up to $10\%$ with $p_\parallel / p_\perp \approx 1.5$. \norm
Like $J_{p_\parallel}$, we observe $J_{nl}$ scales with $\epsilon$,
but the reason is different.
The change in $J_{nl}$ with $\epsilon$ is an artifact:
it is a consequence of preserving $q_0$.

\begin{figure}[!htbp]
\centering
  \includegraphics[width=6.0cm]{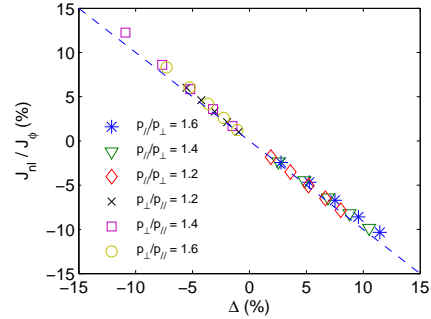}
 \caption{The contribution of nonlinear current $J_{nl}$ to total toroidal current $J_{\varphi}$ in percent as a function of $\Delta$.
  Different markers indicate different magnitude of anisotropy on axis.
 }
 \label{fig:large-delta}
\end{figure}

Figure \ref{fig:large-delta} explores the on-axis contribution of $J_{nl}$ to $J_\varphi$ with changing anisotropy.
It shows that $J_{nl}$ linearly depends on $\Delta$,
but has no dependency on $p_\parallel/p_\perp$, consistent with Eq. (\ref{eq:gs-jphi}).
The result stresses that for analytic working and numerical codes in which $\Delta=0$ approximation is used but anisotropy retained,
care should be taken when anisotropy appears along with $\beta$ above a few percent,
as the effect of this approximation is to delete the nonlinear current.

\begin{figure}[!htb]
$  \begin{array}{c c}
  \includegraphics[width=6cm]{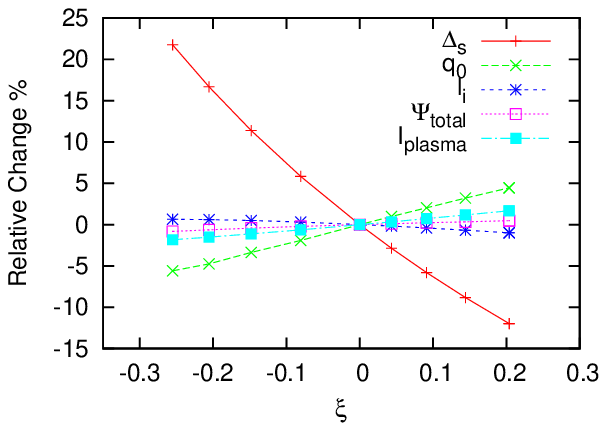} & \includegraphics[width=6cm]{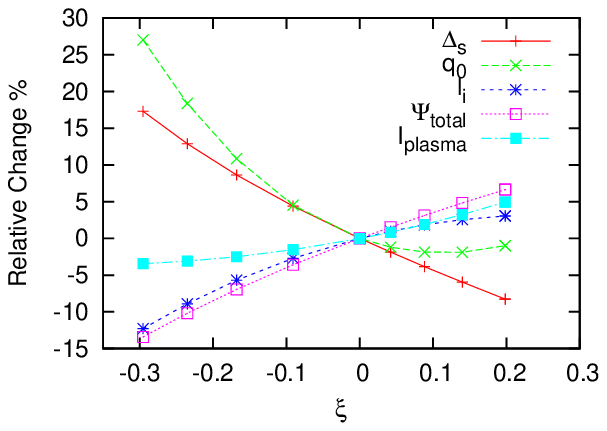}\\
  (a) & (b)
 \end{array}$
   \caption{The change of global parameters: Shafranov-shift($\Delta_s$), $q_0$, $l_i$ (Eq.(\ref{eq:beta-li-def})), total flux, total current
	    due to the changing magnitude of anisotropy based on equilibrium C,
           if the following quantities are hold unchanged for each flux surface:
           (a) $\langle p_\perp \rangle$, $\langle \rho \rangle$ and $\langle RB_\varphi \rangle$,
           (b) $\langle p^*     \rangle$, $\langle \rho \rangle$ and $\langle RB_\varphi \rangle$,
            For instance, the change of $\Delta_s$ is in the form of $(\Delta_{s,aniso}-\Delta_{s,iso})/\Delta_{s,iso}\times 100 \%.$
            }
    \label{fig:bad-p*}
\end{figure}

Inspection of Fig. \ref{fig:jphi-contribution} and \ref{fig:large-delta} shows that at large aspect ratio and low $\Delta$,
$J_\varphi \approx J_{p_\perp} + J_{tf}$. 
Thus, we would expect the global magnetic and current parameters to be insensitive to other changes,
if $p_\perp$ and $RB_\varphi$ profiles remain fixed.
To demonstrate this, we have examined the change in global parameters with changing $\xi = 2(\overline{p_\perp}-\overline{p_\parallel})/(\overline{p_\perp}+\overline{p_\parallel})$, 
but fixed flux surface average profiles $\langle \rho \rangle$, $\langle RB_\varphi \rangle$ and $\langle p_\perp \rangle$ about isotropic equilibrium C,
with ``$\overline{\cdots}$'' the volume average operator and $\langle \cdots \rangle$ the flux surface average operator.
During the scan, we change $\Theta_0$ and adjust $T, H$ and $F$ profiles to keep the above flux surface average profiles identical to equilibrium C.
The percentage change of global parameters is recorded in Fig. \ref{fig:bad-p*}(a),
which shows that with the exception of Shafranov-shift (See Section \ref{sec:sha-shift}),
other global parameters do not change much. 
This confirms the dominant role of $J_{p_\perp} + J_{tf}$ in large aspect ratio tokamaks.
For a comparison, in Fig. \ref{fig:bad-p*}(b), we keep $\langle p^* \rangle$ instead of $\langle p_\perp \rangle$, 
with $p^* = (p_\parallel+p_\perp)/2$ the standard MHD isotropic pressure approximation (See Section \ref{sec:mhd-approximation}). 
As shown in Fig. \ref{fig:bad-p*}(b),
all global parameters will change significantly in the magnitude of $\xi$.
The result shows that $\langle p_\perp \rangle$ is much better than $\langle p^* \rangle$ to retain global parameters,
if $\langle \rho \rangle$ and $\langle RB_\varphi \rangle$ are also unchanged.

\subsection{Deviation from flux function} \label{sec:dev-flux-function}

\subsubsection{Impact on pressure and density} \label{sec:dev-flux-function-profile}

It is clear that with the isotropic assumption $p_\parallel = p_\perp = p$ and static assumption, we have $\nabla p \cdot \bm{B} = 0$,
which means pressure is a flux function.
But now with the additional term $\Delta \bm{BB}$ in Eq. (\ref{eq:basic-gcp-pressure}), 
the two pressures and the density are not flux functions.
This subsection will focus on their variation over a certain flux surface.

If aspect ratio is large, which means the variation of magnetic field on a flux surface, $(B_{max}-B_{min})/B$ is small, 
we can Taylor expand $p_\parallel$ about $B_0 = B(R_0)$, with $R_0$ the major radius of the magnetic axis. 
We use Eq. (\ref{eq:gs-restrictions}) to substitute the partial derivative and derive the difference 
$\Delta p_\parallel \equiv p_{\parallel,out} - p_{\parallel,in}$, 
where the subscript ``out'' denotes the most outward point and ``in'' the most inward point on a flux surface.
Generally $B\approx B_0 R_0 / (R_0 + r \cos{\theta})$ on a flux surface,
in which $r$ is minor radius of a certain flux surface and $\theta$ the poloidal angle. 
Combined, we obtain
\begin{equation}
 \frac{\Delta p_\parallel} { p_\parallel} \approx \frac{2 r}{R_0} \left( \frac{p_\perp - p_\parallel} {p_\parallel} \right)_{R=R_0}.
 \label{eq:ppar-dev-flux-function}
\end{equation}
We note here to reach Eq. (\ref{eq:ppar-dev-flux-function}), 
we don't need any kinetic assumptions.
Similarly, an expansion of $\rho$ and $p_\perp$ about $B_0$, using Eq. (\ref{eq:bi-maxwell-p}), 
(\ref{eq:enthalpy-density-form}) and (\ref{eq:bi-maxwell}),
yields the difference of $\rho$ and $p_\perp$ on a flux surface :
\begin{equation}
 \frac{\Delta \rho} {\rho} \approx 
 \frac{2 r}{R_0} \left( \frac{p_\perp - p_\parallel} {p_\parallel} \right)_{R=R_0}, \   
 \frac{\Delta p_\perp} { p_\perp} \approx 
 \frac{4 r}{R_0} \left( \frac{p_\perp - p_\parallel} {p_\parallel} \right)_{R=R_0},
 \label{eq:dev-flux-function}
\end{equation}
where the meaning of $\Delta \rho$ and $\Delta p_\perp$ is similar to $\Delta p_\parallel$.
Equation (\ref{eq:ppar-dev-flux-function}) and (\ref{eq:dev-flux-function}) indicate the linear dependence
of $\rho$ and $p_{\parallel,\perp}$'s non-flux-function effect on the magnitude of anisotropy and $\epsilon$.
These equations also give the direction of contour shift. 
If $p_\perp > p_\parallel$ ($p_\perp < p_\parallel$), 
the shift of pressures and density contour respect to flux surfaces is outward (inward),
which can be compared to previous findings \cite{cooper-nf-1980,iacono-pfb-1990}.

\begin{figure}[!htb]
\centering
 \includegraphics[width=6.0cm]{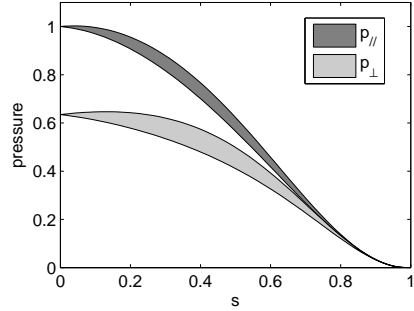}
 \caption{Pressure on flux surfaces for equilibrium A. 
          $s=\sqrt{\Psi/\Psi_0}$ is the standard flux label.
          The upper boundaries of the shaded areas are the maximum value of pressure on certain flux surfaces and the lower boundaries show the minimum.
The shaded areas indicate the range of value on flux surfaces.
           Pressures are normalized to $p_\parallel$ on axis.}
 \label{fig:dev-flux}
\end{figure}

\begin{figure}[!htb]
\centering
  \includegraphics[width=6.0cm]{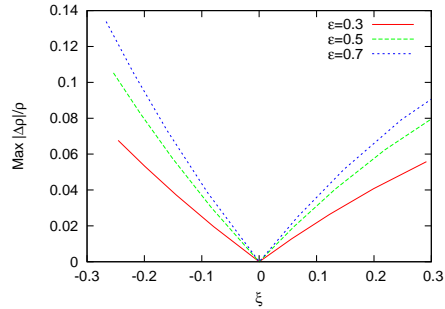}
 \caption{Maximum deviation of $\rho$ from a flux function with different amplitude of anisotropy and different aspect ratio.}
 \label{fig:dev-flux-avg}
\end{figure}

We also study the non-flux function effect numerically.
In Fig. \ref{fig:dev-flux}, 
we plot $p_\parallel$ and $p_\perp$ on different flux surfaces for equilibrium A.
Moving outward from the core, anisotropy decreases and reaches $p_\perp = p_\parallel$ at the boundary,
while $r/R_0$ increases from zero to its maximum at the boundary.
The competition between these two factors makes the difference peak at $s=0.5$,
with $\Delta p_\parallel/p_\parallel \approx 10\%$ and $\Delta p_\perp/p_\perp \approx 20\%$.
This figure demonstrates the deviation of profiles from a function of flux in a single equilibrium.
Figure \ref{fig:dev-flux-avg} shows the maximum in $\Delta \rho / \rho$ as a function of $\epsilon$ and $\xi$,
scanning about the isotropic equilibrium C.
Inspection clarifies the change of density on a flux surface is almost linear with aspect ratio and anisotropy.
Similar behavior is found for $\Delta p_\parallel/ p_\parallel$ and $\Delta p_\perp/ p_\perp$.
Thus, the results of Eq. (\ref{eq:ppar-dev-flux-function}) and (\ref{eq:dev-flux-function}) can be extrapolated to tight aspect ratio tokamaks.


\begin{figure}[!htb]
  \centering
  \includegraphics[width=6.0cm]{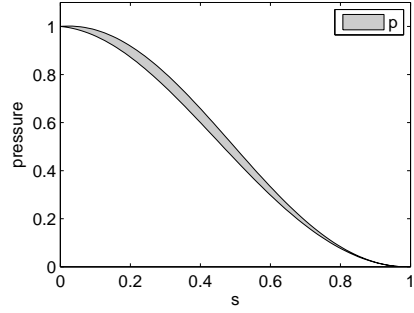}
 \caption{Pressure on flux surfaces for equilibrium D. 
          The upper boundary of the shaded area is the maximum value of pressure on certain flux surfaces and the lower boundary shows the minimum.
The shaded area indicates the range of value on flux surfaces.
           Pressure normalized to 1 at the magnetic axis.}
 \label{fig:intuitive-p}
\end{figure}

To demonstrate the magnitude of the non-flux-function effect, 
we compare the pressure profiles from anisotropic equilibrium A to flowing isotropic equilibrium D.
Figure \ref{fig:intuitive-p} shows the pressure profile on flux surfaces for equilibrium D.
The pressure difference peaks at $7\%$ at $s=0.4$, 
which is comparable to the difference in $p_\parallel$ for static anisotropic equilibrium A.
For equilibrium A, the pressure difference in $p_\perp$ is larger than equilibrium D.

\subsubsection{Impact on Shafranov Shift} \label{sec:sha-shift}
Using methods in \cite{shafranov-ropp-1986, madden-nf-1994, pustovitov-aipcp-2012,goedbloed-am-2010},
for large aspect ratio ($\epsilon = a/R_{0} \ll 1$), low $\beta$ ($\beta\sim\epsilon^{2}$) plasma,
we have to zero's order in $\epsilon$, the modified GSE:
\begin{equation}
 \frac{d}{d\hat{r}} (\mu_0 \langle p_{\perp} \rangle  + \frac{1}{2} B_{\varphi 0}^{2}) + \frac{B_{p0}}{\hat{r}} \frac{d}{d\hat{r}} (\hat{r}B_{p0}) = 0. 
 \label{eq:gs-zero-order-pper}
\end{equation}
Replacing $p_\perp$ by $p$ will return to the original isotropic and static case.
This also confirms our result that flux surface is mostly decided by $p_\perp$ in large aspect ratio scenario.

The next order contribution, $O(\epsilon)$, along with bi-Maxwellian relationships Eq. (\ref{eq:bi-maxwell}),
yields the formulation of Shafranov-Shift: 
\begin{eqnarray}
\fl \Delta_{s}'(\hat{r}) = 
  - \frac{1}{\hat{r} R_{0} B_{p0}^{2}} &\int_{0}^{\hat{r}}\hat{r} d \hat{r} \nonumber\\  
  & \times \left\lbrace  2\hat{r} \mu_0  \left\langle p_\perp 
  \left[1 + \left( \frac{p_\parallel-p_\perp}{2 p_\perp} \right)+ \left( \frac{T_i}{2 T_\perp} M_{t \varphi}^2 \right) \right] \right\rangle'
  -B_{p0}^{2} \right\rbrace.
\label{eq:gs-sha-shift-ani-rot}
\end{eqnarray}
This result is same as \cite{madden-nf-1994,pustovitov-aipcp-2012}.
The variables $B_{p0}$ and $\langle p_{\perp} \rangle$ are related through Eq. (\ref{eq:gs-zero-order-pper}),
and are independent to $p_\parallel$.
Anisotropy and flow contribute to the Shafranov-Shift only through $p_\parallel$ and $M_{t \varphi}^2$,
and their effect is to scale $p_\perp$.
An example of how anisotropy influence Shafranov-shift is provided in Fig. \ref{fig:bad-p*}(a),
where $\langle p_{\perp} \rangle$ and $\langle RB_\varphi \rangle$ are fixed.
The figure shows that $p_\parallel > p_\perp$ ($p_\parallel < p_\perp$) indicates more (less) Shafranov-shift
and the magnitude of this change is linear in $\xi$.

\section{Performance of isotropic model in reconstruction of anisotropic systems} \label{sec:mhd-approximation}
In this section we examine the implications of the choice of model in equilibrium reconstruction. 
A useful starting point are global invariants obtained by integrating momentum conservation. Following this procedure, 
Cooper and Lao \cite{cooper-pp-1982, lao-nf-1985-1} reached the following relationship between global parameters for large aspect ratio tokamaks (Eq.(12) of \cite{lao-nf-1985-1}): 
\begin{equation}
 \frac{1}{2}({\beta}_{p \perp} + {\beta}_{p \parallel}) + W_{pt} + \frac{l_i}{2} = \frac{S_1}{4} + \frac{S_2}{4}(1+\frac{R_t}{R_0}),
 \label{eq:beta-li-restriction}
\end{equation}
with $R_0$ the major radius, $R_t$ a volume dependent constant and
\begin{equation}
  {\beta}_{p \parallel} \equiv \frac{2 \mu_0 \overline{p_{\parallel} }} {B_{pa}^2} ,\ \ 
  {\beta}_{p \perp} \equiv \frac{2 \mu_0 \overline{ p_{\perp} }} {B_{pa}^2} ,\ \ 
  W_{pt} \equiv \frac{\mu_0 \overline{\rho u^2 }} {B_{pa}^2} ,\ \ 
  l_i \equiv \frac {\overline{ B_p^2 }} {B_{pa}^2}, \ \ 
  \label{eq:beta-li-def}
\end{equation}
in which $B_{pa}$ is average poloidal field at boundary and
$u$ is the rotation velocity.
The terms ${\beta}_{p \parallel}$ is the parallel poloidal beta, ${\beta}_{p \perp}$ the perpendicular poloidal beta, 
$W_{pt}$ the rotation poloidal beta and $l_i$ the internal inductance.
In this section, we consider static equilibria in which $W_{pt}=0$.
The constants $S_1, S_2$ are integrals of external fields and currents and therefore can be measured \cite{shafranov-pp-1971}.
For a given set of data from magnetic probes, $S_1$ and $S_2$ are exactly determined.
Equation (\ref{eq:beta-li-restriction}) provides a good measurement of fit for reconstructions using both anisotropic models 
and MHD model with $p = p^*=(p_\parallel+p_\perp)/2$ approximation and $\beta = ({\beta}_{p \perp} + {\beta}_{p \parallel})/2$ (ideal MHD).
This is the historical reason to use $p^*$ as the approximate scalar pressure.
The section intends to answer the question that if plasma is anisotropic and we still reconstruct using ideal MHD,
how good are the reconstructed profiles,
compared to using an anisotropic model.

\subsection{Model dependence in equilibrium reconstuction} \label{sec:mhd-large}

The impact of different models on the inferred pressure and current gradient profiles can be examined by comparison of the toroidal current profile at large aspect ratio. 
For the ideal MHD model, the GSE gives 
\begin{equation}
\mu_0 RJ_{\varphi MHD} = \mu_0 R^2p_{MHD}'(\Psi)+ F_{MHD}F_{MHD}'(\Psi). \label{eq:jphi_GS}
\end{equation}
where we have added the subscript MHD to tag these functions with an ideal MHD model.
A similar functional form can be written for the toroidal current using an anisotropy modified MHD model.
At large aspect ratio, the approximations $R \approx R_0 + r\cos{\theta}$ and $B \approx B_0 R_0 / R$ can be applied.
We also take $\Psi$ derivative on both sides of Eq. (\ref{eq:gs-restrictions}),
and use it to substitute the cross derivative in the Taylor expansion of $\partial p_\parallel/ \partial \Psi$ about $B_0$.
If flow is ignorable, the RHS of the modified GSE Eq. (\ref{eq:gs-2}) can thus be rearranged into
\begin{equation}
  \mu_0 R J_{\varphi m} \approx \mu_0 R^2 p_{0,m}^{*\prime} + \left(F_mF_m' +  \mu_0 R_0^2\frac{p_{\parallel 0 m}' - p_{\perp 0 m}' }{2}\right) + O\left(\frac{r^2}{R^2}\right),  \label{eq:rjphi-regroup}
\end{equation}
where we have similarly added the subscript $m$ to tag the functions with the anisotropy modified MHD model. 
The functions $p_{\parallel 0 m}, p_{\perp 0 m}$ and $p^*_{0 m}$ are those quantities on the flux surface at point $R=R_0$.
Higher order term are written as $O(r^2/R^2)$.

\mody Providing internal current profile information (such as MSE) is available, 
$J_{\varphi MHD} = J_{\varphi m}$ in any reconstruction: the current profile is unique.  
To $O(r/R)$, the RHS of Eq. (40) and (41) have the same variables and functional dependence with $R^2$,
that is, a $R^2$ flux surface varying part and a flux surface invariable part.
By equating these two parts respectively, 
reconstructions using different models but the same data will yield \norm
\begin{equation}
 p_{MHD}' = p_{0,m}^{*\prime}, \label{eq:fake-p}
\end{equation}
\begin{equation}
 F_{MHD}F'_{MHD} = F_m F_m' +  \mu_0 R_0^2\frac{p_{\parallel0m}' - p_{\perp 0m}' }{2} . \label{eq:fake-rbphi}
\end{equation}
Consequently the inferred pressure profile will be identical to the usual $p^*$ approximation, 
but toroidal flux function, and thus the poloidal current profile will be different in the GSE and the modified GSE models. 
This is consistent with Fig. \ref{fig:bad-p*}(b) which shows the plasma cannot preserve its global parameters, 
if we fix both $\langle p^* \rangle$ and $\langle RB_\varphi \rangle$ but vary anisotropy.

At tight aspect ratio,
we should consider $O(r^2/R^2)$ contribution to the modified GSE,
with the second term in Taylor expansion of Eq. (\ref{eq:bi-maxwell-p}), 
(\ref{eq:enthalpy-density-form}) and (\ref{eq:bi-maxwell}).
The result is
\begin{equation}
  f\left(O\left(\frac{r^2}{R^2}\right)\right) = -\mu_0 (p_{\parallel0}-p_{\perp0})\left(1+\frac{p_{\perp0}}{p_{\parallel0}}\right)\frac{r^2}{R_0^2} \cos^2{\theta} + O\left(\frac{r^3}{R^3}\right).
  \label{eq:rjphi-2ndorder}
\end{equation}
Due to the $\cos^2\theta$ dependent, it is not possible to resolve $J_\varphi$ into two MHD flux functions, as done to the zeroth and first parts of Eq. (\ref{eq:rjphi-regroup}).
Equation (\ref{eq:rjphi-2ndorder}) reveals the dependency of the higher order term on the product of $(p_{\parallel0}-p_{\perp0})/p_{\parallel0}$ and $r^2/R^2$.
Thus, in tight aspect ratio tokamaks with large anisotropy, 
the reconstructed $J_\varphi$ and $q$ profile formed by the two flux functions may be distorted,
in comparison to the results from anisotropic reconstruction.

\subsection{Equilibrium reconstruction of a MAST discharge} \label{sec:mhd-tight}

\mody We here study a pair of reconstructions from a single discharge.
The example is from EFIT-TENSOR
reconstruction for MAST($\epsilon\approx0.7$) discharge \#18696 at 290ms,
using either an anisotropic model or isotropic model.
In this discharge, MSE data is not available.
The constraints we used are magnetic probes, total currents and pressures from TRANSP.
These constraints are identical in both reconstructions,
except for the anisotropic reconstruction, 
$p_\parallel$ and $p_\perp$ are constrained to TRANSP $p_\parallel$ and $p_\perp$ respectively,
and for the isotropic reconstruction, isotropic pressure is constrained to $p^* = (p_\parallel + p_\perp)/2$. \norm
\begin{figure}[!htb]
\centering
$ \begin{array}{c c}
  \includegraphics[width=6cm]{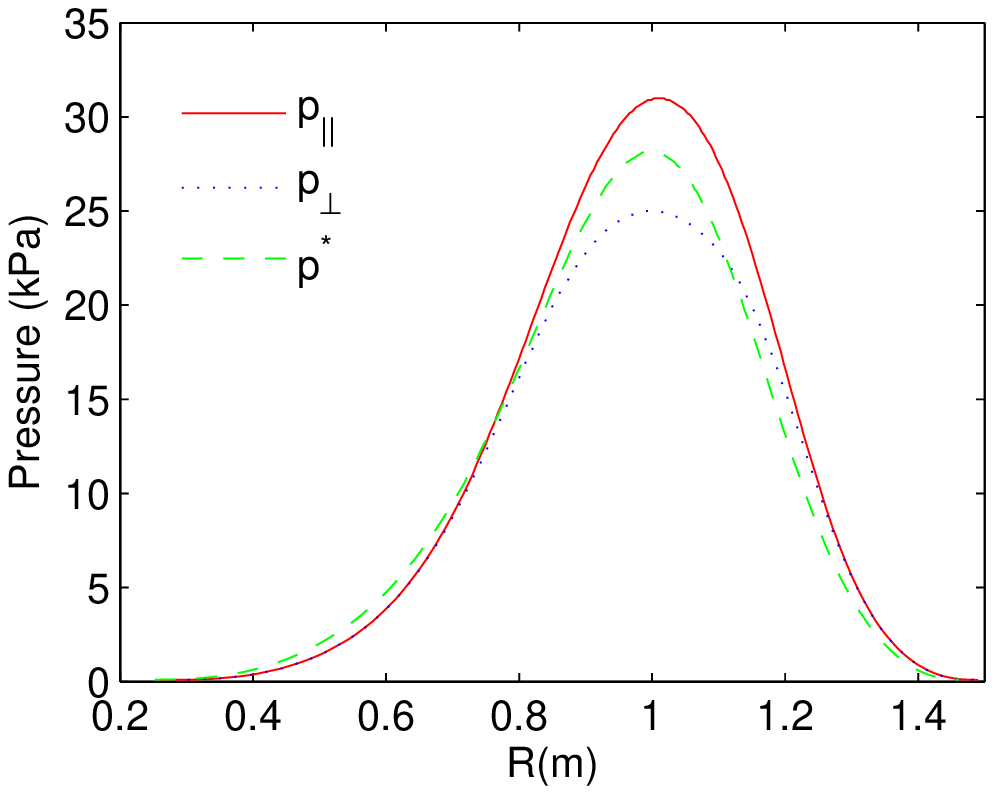} & \includegraphics[width=6cm]{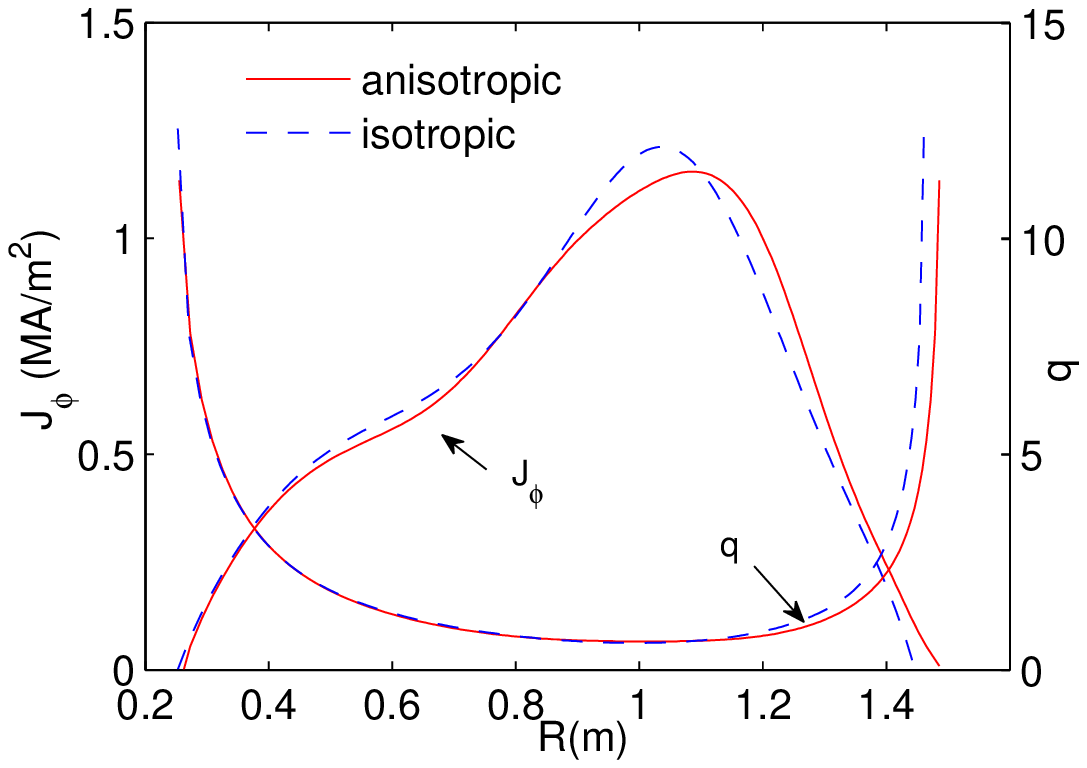} \\
  (a) & (b) \\
  \includegraphics[width=6cm]{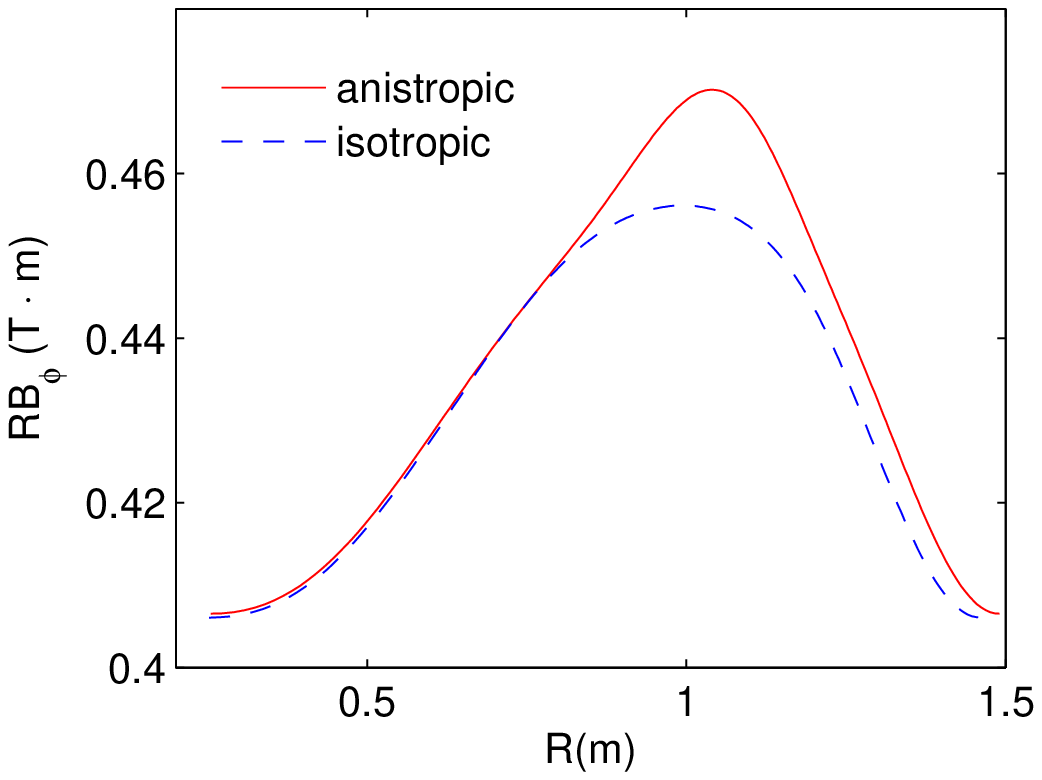} & \includegraphics[width=6cm]{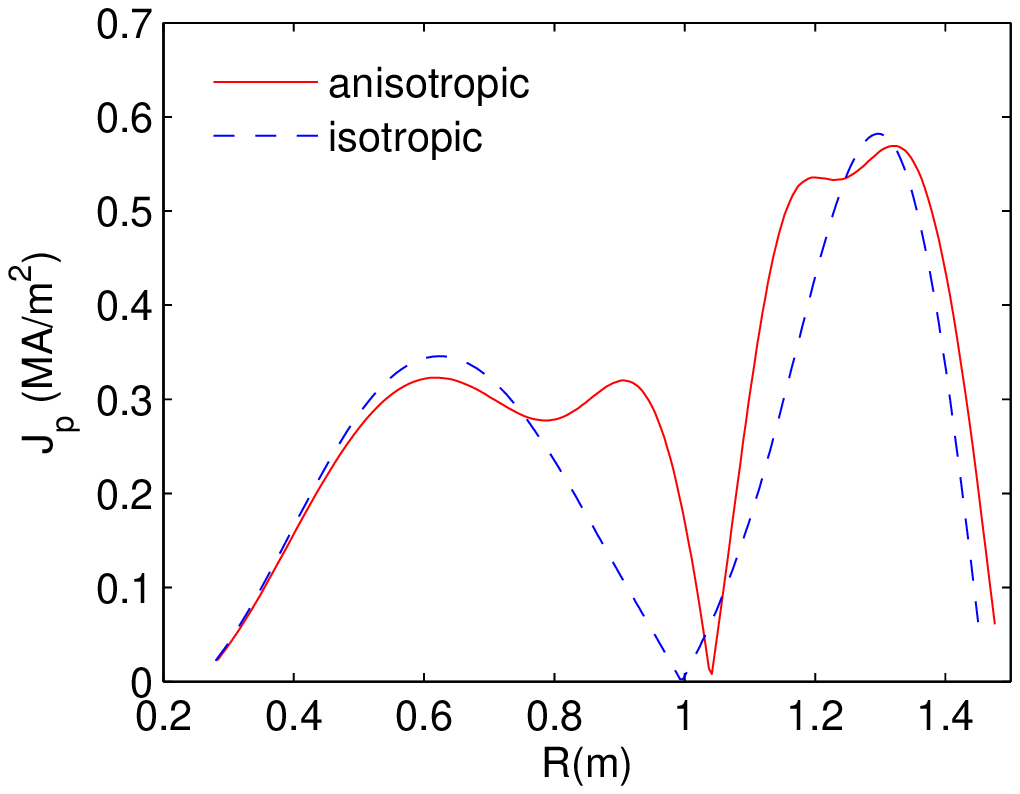} \\
  (c) & (d)
 \end{array}$
 \caption{(a) Pressures on the mid-plane in anisotropic reconstruction (two pressures with solid and dash dot line)
          and in isotropic reconstruction ($p^*$ with dot line) for MAST discharge 18696 at 290ms.
          (b) The reconstructed $J_\varphi$ profile and $q$ profile on the mid-plane. 
          (c) The reconstructed $RB_\varphi$ profile on the mid-plane.
          (d) The reconstructed poloidal current profile on the mid-plane}
 \label{fig:mast-18696}
\end{figure}

In this discharge, NBI is parallel and we have $p_\parallel / p_\perp \approx 1.25$ on the magnetic axis,
as shown in Fig. \ref{fig:mast-18696}(a).
We can see from Fig. \ref{fig:mast-18696}(b) that the two reconstructions gives almost the same $J_\varphi$ profiles,
with a small difference in the core region.
We also notice that these two reconstructions give slightly different boundaries,
causing the difference of $q$ and $J_\varphi$ profile on the low field side.
Both inference differences arise because the EFIT-TENSOR reconstruction is not constrained by a $J_\varphi$ profile.
Despite these differences, the $q$ profile is found to be nearly identical as a function of flux in the two cases.
In our previous work \cite{hole-ppcf-2011}, we recorded a $15\%$ lift in $q_0$ due to anisotropy, which is not observed here.
The reason is that in \cite{hole-ppcf-2011}, 
the two equilibria with/without anisotropy had fixed profiles, 
not fixed external constraints of equilibrium,
as studied here.
In addition, modelled anisotropy in \cite{hole-ppcf-2011} was $p_\perp / p_\parallel = 1.7$ (only the beam pressure was considered).

As predicted by Eq. (\ref{eq:fake-rbphi}),
the MHD reconstructed toroidal field is underestimated in comparison to the anisotropic reconstruction.
This prediction is verified by Fig. \ref{fig:mast-18696}(c),
showing that $RB_\varphi$ is underestimated by 3\% at the core.
When looking at $J_p$ profiles of the two cases in Fig. \ref{fig:mast-18696}(d),
we discover a large discrepancy near the core region,
which peaks at $R=0.9$m with isotropic $J_p$ only $1/3$ of its anisotropic reconstruction.
The difference can be explained by Eq. (\ref{eq:fake-rbphi}).
Since the two models infer different $RB_\varphi'$,
$J_p$ is different through $\mu_0 J_p = |\nabla RB_\varphi|/R$ from Eq. (\ref{eq:jphi}).
In this case
the maximum contribution of the $O({r^2}/{R^2})$ term is $1.5\%$ to the total current,
so the higher order contribution is ignorable.

\subsection{Implications of using MHD to reconstruct anisotropic plasma}
Based on the above findings, if single pressure MHD is used to reconstruct a purely anisotropic plasma,
the following four problems will occur according to aspect ratio and magnitude of anisotropy.

\begin{enumerate}[(i)]
\item The poloidal current is different.

This problem is demonstrated in Section \ref{sec:mhd-large} and \ref{sec:mhd-tight}
and occurs when the variation of $F$ profile is comparable to the variation of $p_\parallel - p_\perp$ across flux surfaces.

\item The anisotropic profiles are not flux functions.

In MHD, $p$, $RB_{\varphi}$ and $\rho$ are flux functions.
As shown in Section \ref{sec:dev-flux-function-profile},
they deviate from flux functions.
According to Eq.(\ref{eq:ppar-dev-flux-function}), (\ref{eq:dev-flux-function}) and Fig. \ref{fig:dev-flux},
this problem linearly increases with $\epsilon$ and $\xi$.

\item Force balance is only satisfied to $O({r}/{R})$ with two flux functions. 

At tight aspect ratio and large anisotropy,
we should take into account terms $O(r^2/R^2)$ in the modified GSE.
It is not possible to decompose the $J_\varphi$ profile into the combination of two flux functions as we demonstrated in Section \ref{sec:mhd-tight}.
If MHD reconstruction is used, the reconstructed $J_\varphi$ profile formed by two flux functions may be distorted.
Inspection of Eq. (\ref{eq:rjphi-2ndorder}) reveals that this problem is a linear function of $\epsilon^2\xi$.

\item The nonlinear current $J_{nl}$ is important at high $\beta$ and large anisotropy.

In Section \ref{sec:flux-surface-shape}, we showed that $J_{nl}$ is proportional to $\Delta$.
The ideal MHD reconstruction neglects $J_{nl}$,
which might impact the accuracy of the reconstructed $J_\varphi$ profile and the $q$ profile in a plasma with high $\beta$ and large anisotropy.
\end{enumerate}

\begin{figure}[!htbp]
\centering
  \includegraphics[width=6cm]{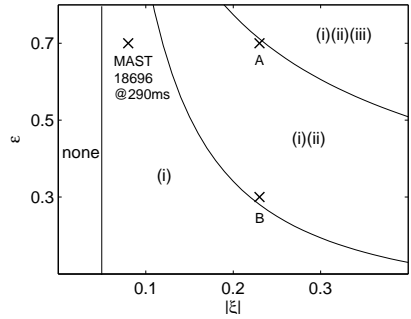}
  \caption{Problems with ideal MHD reconstructions in $\epsilon-\xi$ space.
  The indexes 'i', 'ii' and 'iii' each indicates problem i - iii occur(s) if parameters of an equilibrium is in this region.
  The 'x' markers represent the positions of the MAST \#18696 shot, equilibrium A and B in $\epsilon-\xi$ space, respectively.}
  \label{fig:ani-problem}
\end{figure}

To illustrate the problems in $\epsilon-\xi$ space, we have sketched regimes where each problem might occur.
The corresponding contours are shown in Fig. \ref{fig:ani-problem},
which consist of four regions with a different number of problems.
The lower boundaries are: for problem (i) $|\xi|=0.05$ which represents $5\%$ difference between $p_\parallel$ and $p_\perp$ on average;
for problem (ii) $|\Delta \rho| / \rho = 5\%$ calculated from Fig. \ref{fig:dev-flux-avg},
taking the average of $\xi>0$ and $\xi<0$;
for problem (iii) maximum contribution of the $O(r^2/R^2)$ term to $J_\varphi$ equals to $5\%$ ,
which is obtained by scanning around equilibrium C.
Projection of problem (iv) is not meaningful in $\epsilon-\xi$ space, 
as it is a function of $\Delta$ thus $\beta$ and $\xi$, not $\epsilon$.

\mody We have identified the \#18696 MAST equilibrium and our equilibrium A and B in these contours.
Also, $p_\perp /p_\parallel \approx 2.5$ was found in a JET discharge ($\epsilon \approx 0.3$) during ICRH heating \cite{zwingmann-ppcf-2001}.
The parameter $|\xi|$, if assumed to reach one third of its maximum local value, is $0.3$.
Problem (ii) is significant in this case,
with maximum $\Delta p_\parallel/p_\parallel \approx 17\%$.
Recent unpublished MAST data suggests the existence of discharges with $|\xi|>0.3$,
and thus encounter Problems (i)-(iii).
We will include the study of this discharge in our later publications.
Finally, Problem (iv) appears in discharges with relative high $\beta$.
To date, we haven't identified a discharge with $\Delta > 5\%$ in MAST.
However, a $>40\%$ volume average $\beta$ is observed in NSTX discharges with strong parallel injection \cite{gates-pop-2003}.
Also, the beam power will increase to $7.5MW$ in MAST Upgrade \cite{barrett-fed-2011},
providing possibility to trigger Problem (iv) and to enrich our study in the future. \norm

\section{Conclusion} \label{sec:conclusion}
The impact of pressure anisotropy to plasma equilibrium is studied analytically and numerically.
To achieve the latter,
we have extended the fixed boundary equilibrium and mapping code HELENA to include toroidal flow and anisotropy (HELENA+ATF).
We decompose the toroidal current into contributions from both pressures, the toroidal field and the nonlinear part
and find the dependence of $J_{p_\parallel}$ on the ratio $B_p^2/B^2$.
\mody We find a dominant role of $J_{p_\perp}$ over $J_{p_\parallel}$ in the anisotropy 
and toroidal flow modified Grad-Shafranov equation in large aspect ratio tokamaks.
However in a MAST like equilibrium, the $J_{p_\parallel}$ contribution can reach $20\%$ of the total current 
with $\epsilon=0.7$ and $p_\parallel / p_\perp \approx 1.5$
which should not be ignored.
The impact of this is a $10\%$ change in the current profile,
and thus the $q$ profile, with corresponding implication for plasma stability. \norm
The nonlinear current $J_{nl}$ is proportional to $\Delta$,
and should not be neglected when anisotropy appears in a high $\beta$ plasma.
We have also found that the deviation of profiles from flux functions is in the order of $\epsilon |p_\parallel-p_\perp|/p_\parallel$,
showing a larger contour shift with tighter aspect ratio and larger anisotropy.

Motivated by these analysis, we find that depending upon the aspect ratio and the magnitude of anisotropy,
the following problems may be encountered when the ideal MHD model with $p^*=(p_\parallel+p_\perp)/2$ is used to reconstruct an anisotropic plasma.
First, the poloidal current is different. 
This occurs when the variation of $F$ profile is comparable to the variation of $p_\parallel - p_\perp$ across flux surfaces.
Second, the anisotropy profiles are not flux functions,
their difference on a flux surface linearly increases with the magnitude of anisotropy and $\epsilon$.
Third, the $O(r^2/R^2)$ contribution to $J_\varphi$ is not considered.
This may distort the $J_\varphi$ and $q$ profiles in tight aspect ratio tokamaks with large anisotropy.
Finally, the nonlinear current is neglected, 
degrading the accuracy of the result in a plasma with high $\beta$ and large anisotropy.

\mody In future work, we plan to study the impact of anisotropy on the magnetic configurations,
from a range of experimental discharges and machines,
to address this problem empirically.
We also plan to study the anisotropic effect on plasma stability. \norm

\ack
We gratefully acknowledge the support of G.T.A. Huysmans in providing the HELENA code.
We would also like to thank L. Chang and G. von Nessi for useful discussions in this research.
The project is funded by the China scholarship council, Australian ARC project DP1093797 and FT0991899.

\appendix
\section{Solvability of $p_\parallel, p_\perp, B$ and $\Delta$} \label{app-solve}
Here, we demonstrate that Eq. (\ref{eq:mirror}) and (\ref{eq:firehose}) are a set of sufficient conditions for the four interdependent variables 
$p_\parallel, p_\perp, B$ and $\Delta$ (Eq. (\ref{eq:basic-gcp-pressure}) (\ref{eq:b-definition}) and (\ref{eq:f-definition})) 
to have one and only one root.

The $n$th Picard iteration gives $\Psi_n(R,Z)$ and thus $B_{p,n}=|\nabla\Psi_n|$.
To calculate the magnetic field $B$ after the $n$th iteration at a certain grid point: $B_n(R,Z)$,
the following equations need to be solved for unknown $B_{n}$, with known $\Psi_n, B_{p,n}$ and $R$:
\begin{eqnarray}
 B^2_{n} = \frac{F^2(\Psi_n)}{(1-\Delta_{n})^2 R^2} + B_{p,n}^2, \\  
 \Delta_{n} = \frac{\mu_0 [p_\parallel(\Psi_n, B_{n},R)- p_\perp(\Psi_n, B_{n},R)]}{B_{n}^2}.
 \label{eq:calc-delta}
\end{eqnarray}
Rearranging Eq. (\ref{eq:calc-delta}) and taking the derivative lead to 
\begin{eqnarray}
\label{eq:f-bn+1}
 \fl g(B_{n}) = (B_{n}^2 - B_{p,n}^2)(1-\Delta_{n})^2 -\frac{F^2(\Psi_n)}{R^2}= 0,\\ 
 \fl g'(B_{n}) = 2 B_{n} (1-\Delta_{n})\nonumber\\
     \times \left[(1-\frac{B_{p,n}^2}{B_{n}^2}) \left(1 + \frac{\mu_0}{B_{n}} \frac{\partial p_\perp(\Psi_n, B_{n},R)}{\partial B_{n}} \right)
     + \frac{B_{p,n}^2}{B_{n}^2}(1-\Delta_{n})\right].
\end{eqnarray}
With Eq. (\ref{eq:mirror}), (\ref{eq:firehose}) and $B > B_p$, 
we have $g'(B_{n})>0$. 
Therefore $g(B_{n})$ is monotonically increasing from $B_{p,n}$ to $+\infty$.
Providing that $g(B_{p,n})<0$ and $g(+\infty)\rightarrow +\infty$,
Eq. (\ref{eq:f-bn+1}) should have one and only one root in region $[B_{p,n}, +\infty)$.

\section*{References}
\bibliography{anisotropy_equilibrium}

\begin{thebibliography}{10}

\bibitem{review-nf-2007}
A.~Fasoli, {\it et~al.\/}, {\it Nuclear Fusion\/} {\bf 47}, S264 (2007).

\bibitem{zwingmann-ppcf-2001}
W.~Zwingmann, L.-G. Eriksson, P.~Stubberfield, {\it Plasma Phys. Control.
  Fusion\/} {\bf 43}, 1441 (2001).

\bibitem{hole-ppcf-2011}
M.~Hole, {\it et~al.\/}, {\it Plasma Phys. Control. Fusion\/} {\bf 53}, 074021
  (2011).

\bibitem{northrop-prl-1964}
T.~Northrop, K.~Whiteman, {\it Phys. Rev. Lett.\/} {\bf 12}, 639 (1964).

\bibitem{grad-pf-1967}
H.~Grad, {\it Phys. Fluids\/} {\bf 10}, 137 (1967).

\bibitem{dobrott-pf-1970}
D.~Dobrott, J.~Greene, {\it Phys. Fluids\/} {\bf 13}, 2391 (1970).

\bibitem{spies-pf-1974}
G.~Spies, {\it Phys. Fluids\/} {\bf 17}, 1879 (1974).

\bibitem{salberta-pf-1987}
E.~Salberta, R.~Grimm, J.~Johnson, J.~Manickam, W.~Tang, {\it Phys. Fluids\/}
  {\bf 30}, 2796 (1987).

\bibitem{cooper-nf-1980}
W.~Cooper, G.~Bateman, D.~Nelson, T.~Kammash, {\it Nuclear Fusion\/} {\bf 20},
  985 (1980).

\bibitem{iacono-pfb-1990}
R.~Iacono, A.~Bondeson, F.~Troyon, R.~Gruber, {\it Phys. Fluids B\/} {\bf 2},
  1794 (1990).

\bibitem{pustovitov-ppcf-2010}
V.~Pustovitov, {\it Plasma Phys. Control. Fusion\/} {\bf 52}, 065001 (2010).

\bibitem{shafranov-ropp-1986}
L.~Zakharov, V.~Shafranov, {\it {Reviews of plasma physics}\/} (1986), vol.~11,
  pp. 153--302.

\bibitem{madden-nf-1994}
N.~Madden, R.~Hastie, {\it Nuclear Fusion\/} {\bf 34}, 519 (1994).

\bibitem{pustovitov-aipcp-2012}
V.~Pustovitov, {\it AIP Conf.Proc.\/} pp. 50--64 (2012).

\bibitem{flow-pop-2004}
L.~Guazzotto, R.~Betti, J.~Manickam, S.~Kaye, {\it Phys. Plasmas\/} {\bf 11},
  604 (2004).

\bibitem{hole-ppcf-2013}
M.~Hole, G.~von Nessi, M.~Fitzgerald, {\it Plasma Phys. Control. Fusion\/} {\bf
  55}, 014007 (2013).

\bibitem{grad-guiding-1967}
H.~Grad, {\it {Magneto-Fluid and Plasma Dynamics}\/} (1967), vol.~1, p. 162.

\bibitem{belova-pop-2003}
E.~Belova, N.~Gorelenkov, C.~Cheng, {\it Phys. Plasmas\/} {\bf 10}, 3240
  (2003).

\bibitem{todo-pop-2005}
Y.~Todo, K.~Shinohara, M.~Takechi, M.~Ishikawa, {\it Phys. Plasmas\/} {\bf 12},
  012503 (2005).

\bibitem{fitzgerals-subnf-2013}
M.~Fitzgerald, L.~Appel, M.~Hole, {\it Nuclear Fusion\/} {\bf 53}, 113040
  (2013).

\bibitem{helena-ccp-1991}
G.~Huysmans, J.~Goedbloed, W.~Kerner, {\it {Proc. CP90 Conf. on Computational
  Physics (Amsterdam)}\/} (1991), p. 371.

\bibitem{goedbloed-am-2010}
J.~Goedbloed, R.~Keppens, S.~Poedts, {\it {Advanced Magnetohydrodynamics With
  Applications to Laboratory and Astrophysical Plasma}\/} (Cambridge University
  Press, 2010), pp. 247--306.

\bibitem{tajiri-jpsj-1967}
M.~Tajiri, {\it J. Phys. Soc. Japan\/} {\bf 22}, 1842 (1967).

\bibitem{parker-pr-1958}
E.~Parker, {\it Phys. Rev.\/} {\bf 109}, 1874 (1958).

\bibitem{transp-1979}
R.~Hawryluk, {\it {Physcis of plasmas close to thermonuclear conditions, CEC,
  Brussels}\/} (1979).

\bibitem{mast-ppcf-2003}
R.~Akers, {\it et~al.\/}, {\it Plasma Phys. Control. Fusion\/} {\bf 45}, A175
  (2003).

\bibitem{cooper-pp-1982}
W.~Cooper, A.~Wootton, {\it Plasma Phys. Control. Fusion\/} {\bf 24}, 1183
  (1982).

\bibitem{lao-nf-1985-1}
L.~Lao, H.~St.John, R.~Stambaugh, {\it Nuclear Fusion\/} {\bf 25}, 1421 (1985).

\bibitem{shafranov-pp-1971}
V.~Shafranov, {\it Plasma Phys. Control. Fusion\/} {\bf 13}, 757 (1971).

\bibitem{gates-pop-2003}
D.~a. Gates, {\it Phys. Plasmas\/} {\bf 10}, 1659 (2003).

\bibitem{barrett-fed-2011}
T.~R. Barrett, {\it et~al.\/}, {\it Fusion Eng. Des.\/} {\bf 86}, 789 (2011).

\end{thebibliography}
\end{document}